\documentclass[12pt]{amsart}

\usepackage{color} 
\usepackage{graphicx} 
\usepackage{latexsym,amssymb,amsmath,amsfonts} 
\usepackage{mathrsfs} 
\usepackage{fixltx2e} 

\allowdisplaybreaks[4]
\newtheorem{mylemma}{Lemma}[section]
\newtheorem{mydefinition}{Definition}[section]
\newtheorem{mytheorem}{Theorem}[section]
\newtheorem{mycorollary}{Corollary}[section]
\newtheorem{myexample}{Example}[section]

\definecolor{Light}{gray}{0.85}

\def\abs#1{\left\vert #1 \right\vert}
\def\allpoly{\mbox{$\re\langle X \rangle$}}
\def\allpolyell{\mbox{$\re^{\ell}\langle X \rangle$}}
\def\allpolyx0degn{\mbox{$P_n$}}

\def\allseries{\mbox{$\re\langle\langle X \rangle\rangle$}}

\def\allseriesell{\mbox{$\re^{\ell} \langle\langle X \rangle\rangle$}}

\def\allseriesLC{\mbox{$\re_{LC}\langle\langle X \rangle\rangle$}}
\def\allseriesm{\mbox{$\re^m\langle\langle X \rangle\rangle$}}

\def\allseriesmLC{\mbox{$\re^{m}_{LC}\langle\langle X \rangle\rangle$}}

\def\allseriestildeell{\mbox{$\re^{\tilde{\ell}} \langle\langle X \rangle\rangle$}}

\def\allseriestildemLC{\mbox{$\re^{\tilde{m}}_{LC} \langle\langle X \rangle\rangle$}}

\def\allseriestildeXtildeellLC{\mbox{$\re^{\tilde{\ell}}_{LC} \langle\langle \tilde{X} \rangle\rangle$}}
\def\allseriesellLC{\mbox{$\re^{\ell}_{LC}\langle\langle X \rangle\rangle$}}

\def\bfem#1{{\bf \em #1}} 

\def\bull{\rule{0.08in}{0.08in}} 

\newcommand{\comment}[1]{} 

\def\diag{{\rm diag}}

\def\Endallseries{{\rm End}(\allseries)}

\def\eqref#1{(\ref{#1})} 

\def\liepoly{{\mathcal L}(X)}

\def\lieseries{\widehat{\mathcal L}(X)}

\def\mbf#1{\hbox{\mathversion{bold}$#1$}} 

\def\norm#1{\left\Vert#1\right\Vert}

\def\openbull{\framebox[0.08in][c]{$\;$}} 

\def\re{{\mathbb R}} 

\def\shuffle{{\scriptscriptstyle \;\sqcup \hspace*{-0.05cm}\sqcup\;}}

\def\supp{{\rm supp}}

\def\begals{\[\begin{aligned}}
\def\endals{\end{aligned}\]}
\def\begce{\begin{center}}
\def\endce{\end{center}}
\def\begar{\begin{array}}
\def\endar{\end{array}}
\def\begeq{\begin{equation}}
\def\endeq{\end{equation}}
\def\begdi{\begin{displaymath}}
\def\enddi{\end{displaymath}}
\def\begdis{\begin{eqnarray*}}
\def\enddis{\end{eqnarray*}}
\def\begeqa{\begin{eqnarray}}
\def\endeqa{\end{eqnarray}}
\def\begdes{\begin{description}}
\def\enddes{\end{description}}
\def\begit{\begin{itemize}}
\def\endit{\end{itemize}}
\def\begen{\begin{enumerate}}
\def\enden{\end{enumerate}}
\def\beglar{\left[\begin{array}}
\def\endrar{\end{array}\right]}
\def\begle{\begin{mylemma}}
\def\endle{\end{mylemma}}
\def\begde{\begin{mydefinition}}
\def\endde{\end{mydefinition}}
\def\begth{\begin{mytheorem}}
\def\endth{\end{mytheorem}}
\def\begco{\begin{mycorollary}}
\def\endco{\end{mycorollary}}
\def\begprop{\begin{myproposition}}
\def\endprop{\end{myproposition}}
\def\begex{\begin{myexample}}
\def\endex{\hfill\openbull \end{myexample} \vspace*{0.15in}}
\def\begexer{\begin{myexercise}}
\def\endexer{\end{myexercise}}

\def\begres{\noindent{\bf Remarks}:\begin{enumerate}}
\def\endres{\end{enumerate} \par}
\def\begpr{\noindent{\em Proof:}$\;\;$}
\def\endpr{\hfill\bull \vspace*{0.15in}}
\def\begtab{\begin{tabular}}
\def\endtab{\end{tabular}}
\def\rref#1{(\ref{#1})}

\def\shuff#1#2{\mathbin{
      \hbox{\vbox{\hbox{\vrule \hskip#2 \vrule height#1 width 0pt}\hrule}\vbox{\hbox{\vrule \hskip#2 \vrule height#1 width 0pt\vrule }\hrule}}}}
\def\shuffl{{\mathchoice{\shuff{5pt}{3.5pt}}{\shuff{5pt}{3.5pt}}{\shuff{3pt}{2.6pt}}{\shuff{3pt}{2.6pt}}}}
\def\shuffle{{\, \shuffl \,}}
\def\allseriesLCtensorn{\mbox{$\re_{LC}^{\otimes n}\langle\langle X \rangle\rangle$}}
\def\allseriesLCtensorm{\mbox{$\re_{LC}^{\otimes m}\langle\langle X \rangle\rangle$}}

\begin{document}

\title{Generating Series for Networks of Chen--Fliess Series}

\author{W. Steven Gray}
\address{Department of Electrical and Computer Engineering, Old Dominion University, Norfolk, Virginia 23529, USA}
\email{sgray@odu.edu}
\urladdr{http://www.ece.odu.edu/$\sim$sgray/}
\author{Kurusch Ebrahimi-Fard}
\address{Department of Mathematical Sciences, Norwegian University of Science and Technology (NTNU), 7491 Trondheim, Norway}
\email{kurusch.ebrahimi-fard@ntnu.no}
\urladdr{https://folk.ntnu.no/kurusche/}
\date{}

\begin{abstract} 
Consider a set of single-input, single-output nonlinear systems whose input-output maps are described only in terms of convergent Chen--Fliess series
without any assumption that finite dimensional state space models are available.
It is shown that any additive or multiplicative interconnection of such systems
always has a Chen--Fliess series representation that can be computed explicitly
in terms of iterated formal Lie derivatives.
\end{abstract}


\maketitle

\noindent
\tableofcontents

\thispagestyle{empty}

\section{Introduction}

The study of interconnections of nonlinear control systems is normally posed in a
state space setting. Issues like controllability, observability and
synchronization are natural to consider in this context \cite{Casadei-etal_19,Whalen-etal_15}.
The goal of this paper is to
consider networks of nonlinear systems described only in terms of Chen--Fliess series
without any assumption that finite dimensional state space models are available \cite{Fliess_81,Fliess_83}.
Such models are useful in the context of system identification as relatively few
parameters need to be estimated to yield an accurate approximation of the input-output map \cite{Gray-etal_Autoxx,Venkatesh-etal_CDC19}.
On the other hand, it is not automatically evident that any interconnection of such systems
has a Chen--Fliess series representation. Dynamic output feedback systems, for example, where both
the plant and controller have Chen--Fliess series representations have been shown to always have
such a representation \cite{Gray-etal_SCL14,Gray-Ebrahimi-Fard_SIAM17}.
The proof relies on the contraction mapping theorem applied in
the ultrametric space of noncommutative formal power series. While a perfectly valid approach, it
does not scale easily to complex networks. So in this paper an entirely different approach is taken
based on the notion of a {\em universal control system} due to Kawski and Sussmann \cite{Kawski-Sussmann_97}. The
idea is relatively straightforward in that networks of universal control systems are synthesized leading
to the notion of a {\em formal realization} evolving on an $n$-fold direct product of formal Lie groups. Then the
generating series for any input-output pair in the network is described using the notion of a {\em formal representation}, a
type of infinite dimensional analogue of differential representations that are common
in nonlinear control theory \cite{Isidori_95,Nijmeijer-vanderSchaft_90}. It should be stated, however, that this does {\em not} prove that the resulting Chen--Fliess
series converges in any sense. The tools used here are purely formal and algebraic. As is often the
case when working with Chen--Fliess series, the algebra and the analytic issues can be considered separately with the former providing
the setting for the latter, which is
actually quite convenient \cite{Thitsa-Gray_12}. In particular, it will be shown that any {\em additive} or {\em multiplicative} interconnection of a set of
convergent single-input, single-output Chen--Fliess series always has a Chen--Fliess series representation that can be computed explicitly in terms of iterated formal Lie
derivatives. The problem of determining convergence of the network's generating series will be deferred to future work.

The paper is organized as follows. The next section establishes the notation and terminology of the paper. Section~\ref{sec:formal-realizations}
presents the concept of a formal realization. Formal representations are described in the subsequent section. The main results of the
paper along with several examples are given in Section~\ref{sec:networks-Chen-Fliess-series}. The conclusions are
summarized in the final section, as well as directions for future research.

\section{Preliminaries}

An {\em alphabet} $X=\{ x_0,x_1,$ $\ldots,x_m\}$ is any nonempty and finite set
of noncommuting symbols referred to as {\em
letters}. A {\em word} $\eta=x_{i_1}\cdots x_{i_k}$ is a finite sequence of letters from $X$.
The number of letters in a word $\eta$, written as $\abs{\eta}$, is called its {\em length}.
The empty word, $\emptyset$, is taken to have length zero.
The collection of all words having length $k$ is denoted by
$X^k$. Define $X^\ast=\bigcup_{k\geq 0} X^k$,
which is a monoid under the concatenation (Cauchy) product.
Any mapping $c:X^\ast\rightarrow
\re^\ell$ is called a {\em formal power series}.
Often $c$ is
written as the formal sum $c=\sum_{\eta\in X^\ast}\langle c,\eta\rangle\eta$,
where the {\em coefficient} $\langle c,\eta\rangle\in\re^\ell$ is the image of
$\eta\in X^\ast$ under $c$.
The {\em support} of $c$, $\supp(c)$, is the set of all words having nonzero coefficients.
The set of all noncommutative formal power series over the alphabet $X$ is
denoted by $\allseriesell$. The subset of series with finite support, i.e., polynomials,
is represented by $\allpolyell$.
For any $c,d\in\allseries$, the scalar product is
$\langle c,d\rangle:=\sum_{\eta\in X^\ast} \langle c,\eta\rangle\langle d,\eta\rangle$, provided the sum is finite.
The set $\allseriesell$ is an associative $\re$-algebra under the concatenation product and an associative and commutative $\re$-algebra under
the {\em shuffle product}, that is, the bilinear product uniquely specified by the shuffle product of two words
\begdi
	(x_i\eta)\shuffle(x_j\xi)=x_i(\eta\shuffle(x_j\xi))+x_j((x_i\eta)\shuffle \xi),
\enddi
where $x_i,x_j\in X$, $\eta,\xi\in X^\ast$ and with $\eta\shuffle\emptyset=\emptyset\shuffle\eta=\eta$ \cite{Fliess_81}.
For any letter $x_i\in X$, let $x_i^{-1}$ denote the $\re$-linear {\em left-shift operator} defined by $x_i^{-1}(\eta)=\eta^\prime$
when $\eta=x_i\eta^\prime$ and zero otherwise. It acts as a derivation on the shuffle product.
The Lie bracket $[x_i^{-1},x_j^{-1}]=x_i^{-1}x_j^{-1} - x_j^{-1}x_i^{-1}$ also acts as a derivation on the shuffle product.
Finally, the left-shift operator
is defined
inductively for higher order shifts via $(x_i\eta)^{-1}=\eta^{-1}x_i^{-1}$,
where $\eta\in X^\ast$. For $p\in\allpoly$, let $p^{-1}:=\sum_{\eta\in X^\ast} \langle p,\eta\rangle \eta^{-1}$.

Given any $c\in\allseriesell$ one can associate a causal
$m$-input, $\ell$-output operator, $F_c$, in the following manner.
Let $\mathfrak{p}\ge 1$ and $t_0 < t_1$ be given. For a Lebesgue measurable
function $u: [t_0,t_1] \rightarrow\re^m$, define
$\norm{u}_{\mathfrak{p}}=\max\{\norm{u_i}_{\mathfrak{p}}: \ 1\le
i\le m\}$, where $\norm{u_i}_{\mathfrak{p}}$ is the usual
$L_{\mathfrak{p}}$-norm for a measurable real-valued function,
$u_i$, defined on $[t_0,t_1]$.  Let $L^m_{\mathfrak{p}}[t_0,t_1]$
denote the set of all measurable functions defined on $[t_0,t_1]$
having a finite $\norm{\cdot}_{\mathfrak{p}}$ norm and
$B_{\mathfrak{p}}^m(R)[t_0,t_1]:=\{u\in
L_{\mathfrak{p}}^m[t_0,t_1]:\norm{u}_{\mathfrak{p}}\leq R\}$.
Assume $C[t_0,t_1]$
is the subset of continuous functions in $L_{1}^m[t_0,t_1]$. Define
inductively for each word $\eta=x_i\bar{\eta}\in X^{\ast}$ the map $E_\eta:
L_1^m[t_0, t_1]\rightarrow C[t_0, t_1]$ by setting
$E_\emptyset[u]=1$ and letting
\[E_{x_i\bar{\eta}}[u](t,t_0) =
\int_{t_0}^tu_{i}(\tau)E_{\bar{\eta}}[u](\tau,t_0)\,d\tau, \] where
$x_i\in X$, $\bar{\eta}\in X^{\ast}$, and $u_0=1$. The
{\em Chen--Fliess series} corresponding to $c\in\allseriesell$ is
\begeq
y(t)=F_c[u](t) =
\sum_{\eta\in X^{\ast}} \langle c,\eta\rangle \,E_\eta[u](t,t_0) \label{eq:Fliess-operator-defined}
\endeq
\cite{Fliess_81}.
If there exist real numbers $K_c,M_c>0$ such that
\begdi
\abs{\langle c,\eta\rangle}\le K_c M_c^{|\eta|}|\eta|!,\;\; \forall\eta\in X^{\ast},
\enddi
then $F_c$ constitutes a well defined mapping from
$B_{\mathfrak p}^m(R)[t_0,$ $t_0+T]$ into $B_{\mathfrak
q}^{\ell}(S)[t_0, \, t_0+T]$ for sufficiently small $R,T >0$ and some $S>0$,
where the numbers $\mathfrak{p},\mathfrak{q}\in[1,\infty]$ are
conjugate exponents, i.e., $1/\mathfrak{p}+1/\mathfrak{q}=1$  \cite{Gray-Wang_02}.
(Here, $\abs{z}:=\max_i \abs{z_i}$ when $z\in\re^\ell$.) The set of all such
{\em locally convergent} series is denoted by $\allseriesellLC$, and $F_c$ is
referred to as a {\em Fliess operator}.

Given Fliess operators $F_c$ and $F_d$, where $c,d\in\allseriesellLC$,
the parallel and product connections satisfy $F_c+F_d=F_{c+d}$ and $F_cF_d=F_{c\shuffle d}$,
respectively \cite{Fliess_81}.
When Fliess operators $F_c$ and $F_d$ with $c\in\allseriesellLC$ and
$d\in\allseriesmLC$ are interconnected in a cascade fashion, the composite
system $F_c\circ F_d$ has the
Fliess operator representation $F_{c\circ d}$, where
the {\em composition product} of $c$ and $d$
is given by
\begdi
c\circ d=\sum_{\eta\in X^\ast} \langle c,\eta\rangle\,\psi_d(\eta)(\mathbf{1})
\enddi%
\cite{Ferfera_80}. Here $\mbf{1}$ denotes the monomial $1\emptyset$, and
$\psi_d$ is the continuous (in the ultrametric sense) algebra homomorphism
from $\allseries$ to the vector space endomorphisms on $\allseries$, $\Endallseries$, uniquely specified by
$\psi_d(x_i\eta)=\psi_d(x_i)\circ \psi_d(\eta)$ with
$ 
\psi_d(x_i)(e)=x_0(d_i\shuffle e),
$
$i=0,1,\ldots,m$
for any $e\in\allseries$,
and where $d_i$ is the $i$-th component series of $d$
($d_0:=\mbf{1}$). By definition,
$\psi_d(\emptyset)$ is the identity map on $\allseries$.
It is sometimes useful to associate a unique alphabet with each operator.
For example, let $X=\{x_0,x_1,\ldots,x_m\}$ and
$\tilde{X}=\{\tilde{x}_0,\tilde{x}_1,\ldots,\tilde{x}_{\tilde{m}}\}$.
If $c\in\allseriestildeXtildeellLC$ and
$d\in\allseriestildemLC$, then
the cascade connection $F_c\circ F_d$ has
the generating series in $\allseriestildeell$
\begeq \label{eq:c-circ-d}
c\circ d=\sum_{\tilde{\eta}\in \tilde{X}^\ast} \langle c,\tilde{\eta}\rangle \,\psi_d(\tilde{\eta})(\mbf{1}),
\endeq
where now
$\psi_d(\tilde{x}_i):\allseries\rightarrow \allseries,$ $e\mapsto x_0(d_i\shuffle e)$,
$i=0,1,\ldots,\tilde{m}$. In this case, the letters
in $X$ are identified with the inputs of $F_d$, and the letters
of $\tilde{X}$ are identified with the inputs of $F_c$.
There is a natural isomorphism between $x_0$ and $\tilde{x}_0$ since both
symbols correspond to the unity input ($\tilde{u}_0=u_0=1$).

\begex \label{ex:comp-definition} \rm
Suppose $X=\{x_0,x_1\}$ and $\tilde{X}=\{\tilde{x}_0,\tilde{x}_1\}$.
Let $c=\tilde{x}_1\tilde{x}_1$ and $d=x_1$. The generating series
for the series interconnected system, $c \circ d
=\tilde{x}_1\tilde{x}_1\circ x_1$, can be computed
directly from \rref{eq:c-circ-d} as
\begin{align*}
c \circ d&= \langle c,\tilde{x}_1\tilde{x}_1\rangle \,\psi_d(\tilde{x}_1\tilde{x}_1)(\mbf{1})
= \psi_d(\tilde{x}_1)\circ\psi_d(\tilde{x}_1)(\mbf{1}) \\
&=x_0(x_1\shuffle(x_0(x_1\shuffle \mbf{1})))
=x_0x_1x_0x_1+2x_0x_0x_1x_1.
\end{align*}
It will be shown later (Examples~\ref{ex:comp-formal-realization} and \ref{ex:comp-in-formal-Lie-derivatives})
that this same result can be produced using {\em formal realizations} and {\em formal representations}.
\endex

\section{Formal Realizations}
\label{sec:formal-realizations}

For any finite $T>0$,
$u\in L_1^m[0,T]$ and fixed $t\in[0,T]$, one can associate
the formal power series in $\allseries$
\begdi
	P[u](t)=\sum_{\eta\in X^\ast} \eta\,E_\eta[u](t,0), 
\enddi
which is usually called a {\em Chen series}. If, for example, $u_i(t)=\alpha_i\in\re$, $i=1,2,\ldots,m$ on $[0,T]$ then
$P[u](0)=\mathbf{1}$ and
\begin{align*}
	\frac{d}{dt}P[u](t)&= \sum_{\eta\in X^\ast} \eta \frac{d}{dt}E_{\eta}[u](t,0) \\
	&= \sum_{\eta\in X^\ast}\sum_{i=0}^m  \eta u_i(t)E_{x_i^{-1}(\eta)}[u](t,0) \\
	&=\sum_{\eta\in X^{\ast}} \sum_{i=0}^m  \alpha_ix_i\eta\, E_{\eta}[u](t,0) \\
	&= \left(\sum_{i=0}^m \alpha_i x_i\right) P[u](t).
\end{align*}
It follows directly that
\begdi
\frac{d^n}{dt^n}P[u](0)=\left(\sum_{i=0}^m\alpha_i x_i\right)^n,\;\;n\geq 0,
\enddi
and, therefore
\begdi
	P[u](t)= \sum_{n=0}^\infty \left(\sum_{i=0}^m\alpha_i x_i\right)^n\frac{t^n}{n!}=\exp \left(t\sum_{i=0}^m\alpha_i\, x_i \right).
\enddi
In general, $P[u]$ is the solution to the formal differential equation
\begeq \label{eq:formal-state-equation-in-P}
	\frac{d}{dt}P[u]=\left(\sum_{i=0}^m x_iu_i\right)P[u],\;\;P[u](0)=\mathbf{1},
\endeq
so that $P[u]$ is always the exponential of some Lie element over $X$.
That is, if $\liepoly$ is the free Lie algebra generated by $X$, then
any $d\in\allseries$ is a {\em Lie series} if it can be written in the form
$d=\sum_{n\geq 1} p_n$, where each polynomial $p_n\in\liepoly$ has
support residing in $X^n$.
The set of all Lie series will be denoted by $\lieseries$.
An {\em exponential Lie series} is any series $e=\exp(d):=\sum_{n\geq 0}d^n/n!$,
where $d$ is a Lie series \cite[Chapter 3]{Reutenauer_93}.
In general, \rref{eq:formal-state-equation-in-P} has a solution of the form $P[u](t)=\exp(U(t))$
with $U(t)\in\lieseries$, $t\geq 0$ \cite[Corollary 3.5]{Reutenauer_93}.
As a consequence of
the Baker--Campbell--Hausdorff formula, which states that $\log(\exp({x_i})\exp({x_j}))$ is a Lie series,
the set of all exponential Lie series forms a group, ${\mathcal G}(X)$, under the Cauchy product with unit $\mathbf{1}$
\cite[Lemma 3]{Cartier_56} and \cite[Corollary 3.3]{Reutenauer_93}.

Following the approach of Kawski and Sussmann in \cite{Kawski-Sussmann_97,Sussmann_86}, ${\mathcal G}(X)$ can be viewed as
a {\em formal Lie group} with $\lieseries$ as its corresponding Lie algebra.\footnote{Certain aspects of this framework
can also be found in \cite{Grossman-Larson_92,Grunenfelder_94}.}
A commutative algebra
of real-valued functions on ${\mathcal G}(X)$ is defined using the shuffle algebra on the $\re$-vector space $\allseriesLC$.
Specifically, for any fixed $c\in\allseriesLC$ define $f_c:{\mathcal G}(X)\rightarrow \re$ in terms of the scalar product as
\begeq
\label{eq:comm-real-algebra}
z\mapsto f_c(z)=\langle c,z\rangle=\sum_{\eta\in X^\ast}\langle c,\eta\rangle\langle z,\eta\rangle.
\endeq
Ree's criterion states that $p\in\liepoly$ if and only if $\langle \eta\shuffle\nu,p \rangle=0$ for
all nonempty words $\eta,\nu\in X^\ast$ [Theorem 2.2]\cite{Ree_58}. This implies that $z$ is an exponential Lie
series if and only if $\langle c\shuffle d,z\rangle=\langle c,z\rangle\langle d,z\rangle$ for all $c,d\in\allseries$ \cite[Theorem 3.2]{Reutenauer_93}.
Therefore,
\begdi
	f_{c}(z)f_{d}(z)=\langle c,z\rangle\langle d,z\rangle
	=\langle c\shuffle d,z\rangle=f_{c\shuffle d}(z).
\enddi
Convergence follows from the fact that
the shuffle product is known to preserve local convergence \cite{Wang_90}.\footnote{The authors of \cite{Kawski-Sussmann_97,Sussmann_86}
defined their algebra on $\allpoly$, which entirely avoids the convergence issue. But here $\allseriesLC$ is more suitable for the applications to follow.}
Often $f_c(z)$ will be abbreviated as $c(z)$, which is more natural in the present context.
Analogous to standard Lie group theory, the {\em formal tangent space} at the unit $\mathbf{1}$, $T_{\mathbf{1}}{\mathcal G}(X)$,
is identified with $\lieseries$. Thus, for any fixed $p\in\lieseries$, there is a corresponding tangent vector at $\mbf{1}$
written as the linear functional $V_p(\mbf{1}):\allseriesLC\rightarrow\re$, $c\mapsto V_p(\mbf{1})(c):= \langle c,p\mbf{1}\rangle$ and
satisfying the Leibniz rule\footnote{Recall the definition of the scalar product in the previous section.}
\begin{align*}
V_p(\mbf{1})(c\shuffle d)&=\langle c\shuffle d,p\mathbf{1}\rangle \\
&=\langle p^{-1}(c\shuffle d),\mbf{1}\rangle \\
&=\langle p^{-1}(c)\shuffle d,\mbf{1}\rangle+\langle c\shuffle p^{-1}(d),\mbf{1}\rangle \\
&=\langle p^{-1}(c),\mbf{1}\rangle\langle d,\mbf{1}\rangle+\langle c,\mbf{1}\rangle\langle p^{-1}(d),\mbf{1}\rangle \\
&=\langle c,p\mbf{1}\rangle\langle d,\mbf{1}\rangle+\langle c,\mbf{1}\rangle\langle d,p\mbf{1}\rangle \\
&=V_p(\mbf{1})(c)\, d(\mbf{1})+c(\mbf{1}) V_p(\mbf{1})(d).
\end{align*}
In turn, the tangent space at $z \in {\mathcal G}(X)$, denoted $T_z{\mathcal G}(X)$, is defined via right translation to be the vector space of linear functionals
$V_p(z):\allseriesLC\rightarrow \re$, $c\mapsto V_p(z)(c):=\langle c,pz\rangle$, $p\in\lieseries$ satisfying
\begin{align}
V_p(z)(c\shuffle d)&=\langle c\shuffle d,pz \rangle \nonumber \\
&= \langle c,pz\rangle\langle d,z\rangle+\langle c,z\rangle \langle d,pz \rangle \nonumber \\
&=V_p(z)(c)\,d(z)+c(z) V_p(z)(d). \label{eq:Vpz-derivation}
\end{align}
From a Hopf algebraic viewpoint \cite{Manchon_08}, elements $z \in {\mathcal G}(X)$ are group-like,
that is, for $c,d \in \mathbb{R}_{LC}\langle\langle X \rangle\rangle$ one has $\langle c\shuffle d,z\rangle= \langle c\otimes d,\Delta_\shuffle z\rangle=
\langle c\otimes d,z \otimes z\rangle=\langle  c,z\rangle\langle d,z\rangle$. Here $\Delta_\shuffle$ is the unshuffle
coproduct dualizing the shuffle product. On the other hand, elements $p \in \lieseries$ are primitive, i.e.,
$\Delta_\shuffle p=p \otimes \mbf{1} + \mbf{1} \otimes p$ such that $\langle c\shuffle d,p\rangle =
\langle c,p\rangle\langle d,\mbf{1}\rangle+\langle c,\mbf{1}\rangle\langle d,p\rangle$. Moreover,
$ \Delta_\shuffle pz=  \Delta_\shuffle p \Delta_\shuffle z$ yields $\langle c\shuffle d,pz \rangle=
\langle c,pz\rangle\langle d,z\rangle+\langle c,z\rangle \langle d,pz\rangle$. However, in this work
a Hopf algebraic approach has been suppressed in favor of a purely Lie theoretic presentation.

For any $p\in\lieseries$, the mapping
\begdi
V_p:{\mathcal G}(X)\rightarrow T_z{\mathcal G}(X),\; z\mapsto V_p(z):=pz
\enddi
is a formal right-invariant vector field on ${\mathcal G}(X)$.
Here ${{\mathcal X}}$ will denote the set of all such right-invariant vector fields.
In addition, the {\em formal Lie derivative} is defined to be the mapping
\begdi
	L_p:\allseriesLC\rightarrow\allseriesLC,\;c\mapsto L_pc:=p^{-1}c
\enddi
so that
\begdi
	L_p c(z)=\langle L_pc,z\rangle=\langle p^{-1}c,\;
	z\rangle=\langle c,pz \rangle=V_p(z)(c),
\enddi
and, in particular,
\begin{align*}
L_p (c\shuffle d)(z)
&=\langle L_p (c \shuffle d),z\rangle\\
&=\langle c \shuffle d,pz\rangle\\
&=(L_p c(z))\,d(z)+c(z) L_p d(z),
\end{align*}
which is just an alternative form of \rref{eq:Vpz-derivation}.

Finally, note that \rref{eq:Fliess-operator-defined} can be written
componentwise as
$
y_k(t)=\langle c_k,z(t)\rangle$, $k=1,2,\ldots,\ell,
$
where $c_k\in\allseriesLC$ denotes the $k$-th component of $c\in\allseriesellLC$ and $z(t)=P[u](t)$.
This leads to the following definition.

\begde \label{def:formal-realization-Fc}
For any $c\in\allseriesellLC$, the
\bfem{formal realization} of the Fliess operator $y=F_c[u]$ is
\begin{align*} 
\dot{z}&=\sum_{i=0}^m x_izu_i,\;\; z(0)={\mbf 1} \\
y_k&=\langle c_k,z\rangle,\;\; k=1,2,\ldots,\ell.
\end{align*}
\endde

Observe that
\begin{align*}
L_{x_i}c_k(\mbf{1})&=x_i^{-1}c_k(\mbf{1})=\langle x_i^{-1}c_k,\mbf{1}\rangle=\langle c_k,x_i\rangle \\
L_{x_j}L_{x_i}c_k(\mbf{1})&=x_j^{-1}x_i^{-1}c_k(\mbf{1})=\langle x_j^{-1}x_i^{-1}c_k,\mbf{1}\rangle
=\langle c_k,x_ix_j\rangle,
\end{align*}%
so that the coefficients of $c_k$ can always be written in terms of formal Lie derivatives as
\begin{align}
\label{eq:formal-representation_n_equal_1}
\langle c_k,\eta\rangle&=\langle c_k,x_{i_1}\cdots x_{i_k}\rangle \nonumber \\
&=L_{x_{i_k}}\cdots L_{x_{i_1}}c_k(\mbf{1})=:L_{\eta}c_k(\mbf{1}).
\end{align}

The notion of a formal realization in Definition~\ref{def:formal-realization-Fc} is now extended by taking a finite number of direct products of ${\mathcal G}(X)$, i.e.,
${\mathcal G}^n(X):={\mathcal G}(X)\times {\mathcal G}(X)\times\cdots\times {\mathcal G}(X)$, where ${\mathcal G}(X)$ appears $n$ times.
For any $\hat{c}=c_1\otimes\cdots\otimes c_n\in\allseriesLCtensorn$ define
\begin{align*}
f_{\hat{c}}&:{\mathcal G}^n(X)\rightarrow \re \\
&z\mapsto (c_1\otimes\cdots\otimes c_n)(z_1,\ldots,z_n)=\langle c_1,z_1\rangle\cdots\langle c_n,z_n\rangle.
\end{align*}
A commutative algebra on the $\re$-vector space of all such
real-valued functions on ${\mathcal G}^n(X)$ is given by defining
\begin{align*}
f_{\hat{c}}(z)f_{\hat{d}}(z)&=[\langle c_1,z_1\rangle\cdots\langle c_n,z_n\rangle][\langle d_1,z_1\rangle\cdots\langle d_n,z_n\rangle] \\
&=\langle c_1\shuffle d_1,z_1\rangle\cdots \langle c_n\shuffle d_n,z_n\rangle \\
&=:(\hat{c}\shuffle \hat{d})(z_1,z_2,\ldots,z_n) \\
&=f_{\hat{c}\shuffle \hat{d}}(z).
\end{align*}
As earlier, $f_{\hat{c}}(z)$ will often be abbreviated as $\hat{c}(z)$.
The Lie algebra of
${\mathcal G}^n(X)$, denoted by $\widehat{{\mathcal L}}^n(X)$, is similarly defined as the $n$-fold direct sum of the Lie algebra
${\widehat{\mathcal L}}(X)$ for ${\mathcal G}(X)$ with itself.
The {\em formal tangent space} at the unit $\mathbf{1}_n:=(\mathbf{1},\ldots,\mathbf{1})$, $T_{\mathbf{1}_n}{\mathcal G}^n(X)$,
is identified with $\widehat{{\mathcal L}}^n(X)$
via the one-parameter subgroup $H(t):=(\exp(t p_1),\exp(t p_2),\ldots,\exp(t p_n))$, $p=(p_1,p_2,\ldots,p_n)\in\widehat{{\mathcal L}}^n(X)$
so that $\dot{H}(0)=p$.
For any fixed $p\in\widehat{{\mathcal L}}^n(X)$,
there is a corresponding tangent vector at $\mbf{1}_n$
represented by the linear functional
\begdi
V_p(\mbf{1}_n):\allseriesLCtensorn\rightarrow\re,\; \hat{c}\mapsto \frac{d}{dt}(\hat{c}\circ H(t))|_{t=0}.
\enddi
Observe that
\begin{align*}
V_p(\mbf{1}_n)(\hat{c})&=\frac{d}{dt}\big(\langle c_1,\exp(t p_1)\rangle\cdots\langle c_i,\exp(t p_i)\rangle\cdots\langle c_n,\exp(t p_n)\rangle\big) |_{t=0} \\
&=\sum_{i=1}^n \langle c_1,\mbf{1}\rangle \cdots \langle c_i,p_i\mathbf{1}\rangle\cdots\langle c_n,\mbf{1}\rangle
\end{align*}
satisfies the Leibniz rule:
\begin{align*}
V_p(\mbf{1}_n)(\hat{c}\shuffle \hat{d})
&=\sum_{i=1}^n \langle c_1\shuffle d_1,\mbf{1}\rangle \cdots \langle c_i\shuffle d_i,p_i\mathbf{1}\rangle\cdots\langle c_n\shuffle d_n,\mbf{1}\rangle \\
&=\sum_{i=1}^n \langle c_1\shuffle d_1,\mbf{1}\rangle \cdots \langle p_i^{-1}(c_i\shuffle d_i),\mathbf{1}\rangle\cdots\langle c_n\shuffle d_n,\mbf{1}\rangle \\
&=\sum_{i=1}^n \langle c_1\shuffle d_1,\mbf{1}\rangle \cdots \langle p_i^{-1}(c_i)\shuffle d_i,\mathbf{1}\rangle\cdots\langle c_n\shuffle d_n,\mbf{1}\rangle+ \\
&\hspace*{0.2in}\sum_{i=1}^n \langle c_1\shuffle d_1,\mbf{1}\rangle \cdots \langle c_i\shuffle p_i^{-1}(d_i)),\mathbf{1}\rangle\cdots\langle c_n\shuffle d_n,\mbf{1}\rangle \\
&=V_p(\mbf{1}_n)(\hat{c})\hat{d}(\mathbf{1}_n)+ \hat{c}(\mbf{1}_n) V_p(\mbf{1}_n)(\hat{d}).
\end{align*}
The tangent space at $z \in {\mathcal G}^n(X)$, denoted $T_z{\mathcal G}^n(X)$, is defined via right translation to be the vector space of linear functionals
\begin{align*}
V_p(z)&:\allseriesLCtensorn\rightarrow\re \\
&\hat{c}\mapsto \sum_{i=1}^n \langle c_1,z_1\rangle \cdots \langle c_i,p_iz_i\rangle\cdots\langle c_n,z_n\rangle
\end{align*}
so as to satisfy
\begdi
V_p(z)(\hat{c}\shuffle \hat{d})=V_p(z)(\hat{c})\,\hat{d}(z)+\hat{c}(z) V_p(z)(\hat{d}).
\enddi
For any $p\in\widehat{{\mathcal L}}^n(X)$, the mapping
\begdi
V_p:{\mathcal G}^n(X)\rightarrow T_z{\mathcal G}^n(X),\; z\mapsto (p_1z_1,\ldots,p_nz_n)
\enddi
is a formal right-invariant vector field on ${\mathcal G}^n(X)$.
Here ${\mathcal X}^n$ will denote the set of all such right-invariant vector fields.
In this context, the formal Lie derivative is defined to be the mapping
\begin{align*}
L_p&:\allseriesLCtensorn\rightarrow\allseriesLCtensorn \\
&c_1\otimes\cdots\otimes c_n\mapsto \sum_{i=1}^n c_1\otimes\cdots\otimes p_i^{-1}(c_i)\otimes\cdots\otimes c_n
\end{align*}
so that
\begin{align}
L_p \hat{c}(z)&=\left(\sum_{i=1}^n c_1\otimes\cdots\otimes p_i^{-1}(c_i)\otimes\cdots\otimes c_n\right) (z_1,\ldots,z_n) \nonumber\\
&=\sum_{i=1}^n \langle c_1,z_1\rangle \cdots \langle c_i,p_iz_i\rangle\cdots\langle c_n,z_n\rangle \nonumber \\
&=V_p(z)(\hat{c}), \label{eq:Lpcz-equals-Vpzc}
\end{align}
and directly
\begdi
L_p (\hat{c}\shuffle \hat{d})(z)=(L_p \hat{c}(z))\,\hat{d}(z)+\hat{c}(z) L_p \hat{d}(z).
\enddi

In this generalized setting, a set of $n$ systems with state
$z=(z_1,z_2,\ldots,z_n)$
evolves on the group ${\mathcal G}^n(X)$ according to the
formal state equations
\begdi
\dot{z}_j=\sum_{i=0}^m x_iz_j u_{ij},\;\;z_j(0)=\mbf{1},
\enddi
where $u_{ij}\in L_\mathfrak{p}[0,T]$ and $u_{0j}=1$ for $i=1,2,\ldots,m$, $j=1,2,\ldots,n$. Define $\ell$ outputs
$y_{k}=\hat{c}_{k}(z)$, where $\hat{c}_{k}\in\allseriesLCtensorn$, $k=1,2,\ldots,\ell$.
Therefore, the corresponding input-output map $u\mapsto y$ takes an $m\times n$ matrix of inputs to
$\ell$ outputs. Consider now the situation where a network is formed by allowing each system input to be interconnected to
some function of other systems' outputs and a new external input $v_{ij}$ to yield a new input-output map $v\mapsto y$,
for example, $u_{ij}=\hat{d}_{ij}(z)+v_{ij}$, where
$\hat{d}_{ij}\in\allseriesLCtensorn$. In this case, the state equations for the interconnected system become
\begdi
\dot{z}_j=x_0z_j+\sum_{i=1}^m x_i\hat{d}_{ij}(z)z_j+ x_iz_j v_{ij},\;\;z_j(0)=\mbf{1}.
\enddi
Note, in particular, the appearance of state dependent vector fields $p_jz_j$ with $p_j(t)=\sum_{i=1}^m x_i\hat{d}_{ij}(z(t))\in\lieseries$.
The solution
to $\dot{z}_j=p_jz_j$, $z_j(0)=\mbf{1}$ has the form $z_j(t)=\exp(U_j(t))$, where $U_j(t)\in \widehat{{\mathcal L}}(X)$.
The corresponding
tangent vector at $z(t)$ is
\begin{align}
V_{p(t)}(z(t))&:\allseriesLCtensorn\rightarrow\re \nonumber \\
&\hat{c}\mapsto \frac{d}{dt}(\hat{c}\circ z(t)) \nonumber \\
&=\sum_{j=1}^n \langle c_1,z_1(t)\rangle \cdots \langle c_j,p_j(t)z_j(t)\rangle\cdots\langle c_n,z_n(t)\rangle \nonumber \\
&=L_{p(t)}\hat{c}(z(t)). \label{eq:Lpthatc}
\end{align}
Substituting $p_j(t)=\sum_{i=1}^m x_i\hat{d}_{ij}(z(t))$ on the right-hand side above, where
$\hat{d}_{ij}(z(t))=$\linebreak
$\langle d^{(1)}_{ij},z_1(t)\rangle\cdots\langle d^{(n)}_{ij},z_n(t)\rangle$,
gives
\begin{align}
L_{p(t)}\hat{c}(z(t)) &= \sum_{j=1}^n \langle c_1,z_1(t)\rangle \cdots \langle c_j,p_j(t)z_j(t)\rangle\cdots\langle c_n,z_n(t)\rangle \nonumber \\
&=\sum_{j=1}^n \langle c_1,z_1(t)\rangle \cdots \sum_{i=1}^m \hat{d}_{ij}(z(t))\langle c_j,x_iz_j(t)\rangle \cdots \langle c_n,z_n(t)\rangle \phantom{mm} \nonumber \\
&=\sum_{i=1}^m\sum_{j=1}^n \langle c_1\shuffle d^{(1)}_{ij},z_1(t)\rangle\cdots \langle x_i^{-1}(c_j\shuffle d^{(j)}_{ij}),z_j(t)\rangle
\cdots\langle c_n\shuffle d^{(n)}_{ij},z_n(t)\rangle \nonumber \\
&=:\hat{c}'(z(t)). \label{eq:c-prime-z}
\end{align}
In this way, a second Lie derivative can now be computed directly using \rref{eq:Lpthatc}, thus circumventing the difficult task of explicitly composing time-varying vector fields. Henceforth, all such state dependent Lie series will be written as $p(z)$. No other type of
state dependent series will appear in this paper.
In this context, a generalization of Definition~\ref{def:formal-realization-Fc}
is presented.

\begde
Let $V_i\in {\mathcal X}^n$, $i=0,1,\ldots,m$ with
\begin{align*}
V_i&:{\mathcal G}^n(X)\rightarrow T_z{\mathcal G}^n(X) \\
&z=(z_1,\ldots,z_n)\mapsto V_i(z)=(V_{i1}(z)z_1,\ldots,V_{in}(z)z_n),
\end{align*}
where $V_{ij}(z(t))\in \widehat{\mathcal L}(X)$.
The $j$-th component of the corresponding state equation on ${\mathcal G}^n(X)$ is
\begeq \label{eq:formal-state-equation-generalized}
\dot{z}_j=\sum_{i=0}^m V_{ij}(z)z_j u_{ij} , \;\; z_j(0)=z_{j0}.
\endeq
Given $\hat{c}_{k}\in\allseriesLCtensorn$, $k=1,2,\ldots,\ell$, the $k$-th output equation is defined to be
\begeq \label{eq:formal-output-equation-generalized}
y_{k}=\hat{c}_{k}(z).
\endeq
Collectively, $(V,z_0,\hat{c})$ is a \bfem{formal realization} on ${\mathcal G}^n(X)$ of the formal input-output map
$u\mapsto y$.
\endde

For convenience the integer $n$ will be referred to here as the {\em dimension} of the realization, though this is a misnomer as
the underlying group ${\mathcal G}(X)$ is {\em not} finite dimensional, therefore neither
is the state $z$. The following example illustrates how the concept
naturally arises when Chen--Fliess series are composed.

\begex \label{ex:comp-formal-realization} \rm
Reconsider the systems $y_2=F_{c}[u_2]$ and $y_1=F_{d}[u_1]$ in Example~\ref{ex:comp-definition}
using the same alphabet $X=\{x_0,x_1\}$ for both series.
Each has a formal realization of the form given in Definition~\ref{def:formal-realization-Fc}. Setting
$u_2=y_1$ so that
$y_2=F_{c}\circ F_{d}[u_1]$ yields a formal realization of dimension two:
\begin{align*}
\dot{z}_1&=x_0z_1+x_1z_1u_1,\;\; z_1(0)=\mbf{1} \\
\dot{z}_2&=(x_0+x_1\langle d,z_1\rangle)z_2,\;\; z_2(0)=\mbf{1} \\
y_2&=\langle\mbf{1},z_1\rangle\langle c,z_2 \rangle.
\end{align*}
(Note that $\langle\mbf{1},z_1\rangle=1$.)
Therefore,
\begdi
V_0(z)=
\left[
\begin{array}{c}
x_0z_1 \\
(x_0+x_1\langle d,z_1\rangle)z_2
\end{array}
\right],
\;\;
V_1(z)=
\left[
\begin{array}{c}
x_1z_1 \\
0
\end{array}
\right],
\enddi
and $\hat{c}=\mbf{1}\otimes c$. Observe that the composition $F_{c}\circ F_{d}=F_{c \circ d}$ introduces in the second component of the tangent vector $V_0(z)$ a $z_1$ dependence. The aim is to express $c\circ d$ directly in terms of $(V,\mbf{1}_2,\hat{c})$. This leads to the notion of a {\em formal representation} of a series
as presented in the next section. It can be viewed as a generalization of \rref{eq:formal-representation_n_equal_1}.
\endex

\section{Formal Representations}

The following definition is a formal analog of a differential representation as appears, for example, in \cite{Isidori_95,Nijmeijer-vanderSchaft_90}.

\begde \label{def:differential_representation}
A \bfem{formal representation} of a series $d\in\allseries$ is any triple
$(\mu,z_0,\hat{c})$, where
\begdi
\mu:X^{\ast}\rightarrow {\mathcal X}^n,\; x_i\mapsto V_i
\enddi
defines a monoid homomorphism, $z_0\in{\mathcal G}^n(X)$, and
$\hat{c}\in\allseriesLCtensorn$, so that for any word $\eta=x_{i_k}x_{i_{k-1}}\cdots x_{i_1}\in X^\ast$
\begeq \label{eq:formal-representation}
\langle d,\eta\rangle=L_{\mu(\eta)}\hat{c}(z_0):=L_{\mu(x_{i_1})}L_{\mu(x_{i_2})}\cdots L_{\mu(x_{i_k})}\hat{c}(z_0).
\endeq
By definition, $\langle d,\emptyset \rangle=L_{\emptyset}\hat{c}(z_0):=\hat{c}(z_0)$.
The integer $n\geq 1$ will be called the \bfem{dimension} of the representation.
\endde

\begex \rm
For the trivial case where $n=1$, $\mu(x_i)=x_i$, $z_0=\mbf{1}$, and $d=\hat{c}=c$ it is immediate that \rref{eq:formal-representation} reduces
to \rref{eq:formal-representation_n_equal_1} with $\ell=1$.
\endex

The following lemma provides a sufficient condition under which formal representations are always well defined.

\begle \label{lem:well-defined-formal-representations}
Given $(\mu,z_0,\hat{c})$, if for each $x_i\in X$ $[\mu(x_i)]_j(z):=V_{ij}(z)z_{j}$ with $V_{ij}(z)$ being
some Lie polynomial in ${\mathcal L}(X)$, then
there exists a well defined $d\in\allseries$ satisfying \rref{eq:formal-representation}.
\endle

\begpr
If $(\mu,z_0,\hat{c})$ is a formal representation of $d$ then necessarily for any $\eta=x_{i_1}\cdots x_{i_k}\in X^\ast$
\begdi
\langle d,x_{i_k}\cdots x_{i_1} \rangle=L_{\mu(x_{i_1})}L_{\mu(x_{i_2})}\cdots L_{\mu(x_{i_k})}\hat{c}(z_0),
\enddi
where each $V_{ij}(z)$ is assumed to be a Lie polynomial. Therefore, each Lie derivative can be written as a
polynomial in functions of the form $\langle e,p_iz_i\rangle$ with $p_i\in\liepoly$, $i=1,2,\ldots,n$, and $e\in\allseriesLC$,
implying that
$d$ is well defined, in fact, locally finite \cite{Berstel-Reutenauer_88}.
\endpr

\begex \label{ex:comp-in-formal-Lie-derivatives} \rm
Continuing Examples~\ref{ex:comp-definition} and \ref{ex:comp-formal-realization},
the claim is that $c\circ d$ has a formal representation $(\mu,\mbf{1}_2,\hat{c})$, where
$\mu$ is defined in terms of the vector fields $V_0$ and $V_1$ in Example~\ref{ex:comp-formal-realization} and
$\hat{c}=\mbf{1}\otimes c$. Note that both vector fields satisfy the condition in Lemma~\ref{lem:well-defined-formal-representations}.
As an example, it is verified that
\begdi
	\langle x_0^2x_1^2,c \circ d\rangle
	=L_{\mu(x_0^2x_1^2)}\hat{c}(\mbf{1})
	=L_{V_1}L_{V_1}L_{V_0}L_{V_0}\hat{c}(\mbf{1})=2.
\enddi
First apply \rref{eq:Lpthatc} (suppressing all $t$ dependence)
\begin{align*}
	L_{V_0}\hat{c}(z)
	&=\langle c,V_{02}(z)z_2\rangle \\
	&=\langle x_1^2,(x_0+x_1\langle x_1,z_1\rangle)z_2\rangle.
\end{align*}
Regarding the $z_1$ dependence of $V_{02}(z)$, use \rref{eq:c-prime-z} to get
\begdi
	L_{V_0}\hat{c}(z)
	=\langle x_1,z_1 \rangle \langle x_1,z_2 \rangle
	= (x_1 \otimes x_1)(z_1,z_2)= \hat{c}'(z).
\enddi
Applying \rref{eq:Lpthatc} and \rref{eq:c-prime-z} a second time gives:
\begin{align*}
L_{V_0}L_{V_0}\hat{c}(z)&=L_{V_0}\hat{c}'(z)\\
&=\langle x_1, V_{01}(z)z_1 \rangle \langle x_1,z_2 \rangle
+ \langle x_1,z_1 \rangle \langle x_1,V_{02}(z)z_2 \rangle \\
&= \langle x_1,x_0z_1\rangle\langle x_1,z_2\rangle
+ \langle x_1,z_1\rangle \langle x_1,(x_0+x_1\langle x_1,z_1\rangle)z_2\rangle\\
&=\langle x_1,z_1 \rangle^2 \langle \mbf{1},z_2 \rangle \\
&=\langle x_1 \shuffle x_1 ,z_1 \rangle \langle \mbf{1},z_2 \rangle \\
&=(x_1 \shuffle x_1 \otimes \mbf{1})(z_1,z_2) \\
&=( 2x_1^2\otimes \mbf{1}) (z_1,z_2) = \hat{c}''(z).
\end{align*}
Continuing in this fashion,
\begin{align*}
L_{V_1}L_{V_0}L_{V_0}\hat{c}(z)
&= L_{V_1}\hat{c}''(z)
=\langle 2x_1,z_1 \rangle \langle \mbf{1},z_2 \rangle\\
&=( 2x_1\otimes \mbf{1} )( z_1,z_2) = \hat{c}'''(z)
\end{align*}
and
\begin{align*}
L_{V_1}L_{V_1}L_{V_0}L_{V_0}\hat{c}(z)
&= L_{V_1}\hat{c}'''(z)=\langle 2\mbf{1},z_1 \rangle \langle \mbf{1},z_2 \rangle.
\end{align*}
Therefore, $\langle x_0^2x_1^2,c\circ d\rangle=L_{V_1}L_{V_1}L_{V_0}L_{V_0}\hat{c}(\mbf{1})=2$ as anticipated.
\endex

The proposition in the previous example is established in the general case by the following theorem.

\begth \label{th:generating-series-from-formal-representations}
If $d\in\allseries$ has a well defined formal representation $(\mu,z_0,\hat{c}_{k})$, then the
input-output map $u\mapsto y_{k}$ of the corresponding formal realization
\rref{eq:formal-state-equation-generalized}-\rref{eq:formal-output-equation-generalized}
has a Chen--Fliess series representation with generating series~$d$.
\endth

\begpr
Without loss of generality, assume there is a single output so that the subscripts on $\hat{c}_k$ and $y_k$ can be dropped.
Likewise, assume $n=1$ so the index on the state can be omitted.
Since $\dot{z}(t)$ is a tangent vector at $z(t)\in {\mathcal G}(X)$ for any $t\geq0$, it follows directly
from \rref{eq:Lpcz-equals-Vpzc} that
\begin{align*}
\dot{z}(t)(\hat{c})&=\sum_{i=0}^m V_i(z(t))(\hat{c})u_i(t) \\
&=\sum_{i=0}^m L_{V_i}\hat{c}(z(t))u_i(t).
\end{align*}
Integrating both sides on $[0,t]$ and applying \rref{eq:c-prime-z} gives
\begin{align}
\hat{c}(z(t))&=\hat{c}(z_0)+\sum_{i=0}^m \int_0^t L_{V_i}\hat{c}(z(\tau))u_i(\tau)\,d\tau \nonumber \\
&=\hat{c}(z_0)+\sum_{i=0}^m \int_0^t \hat{c}_i^{\prime}(z(\tau))u_i(\tau)\,d\tau, \label{eq:c-prime-z-again}
\end{align}
where $ L_{V_i}\hat{c}(z(\tau))=\hat{c}_i^{\prime}(z(\tau))=\langle \hat{c}_i^\prime,z(\tau)\rangle$.
Substituting $\hat{c}_i^\prime$ for $\hat{c}$ above yields
\begeq \label{eq:hat-c-prime-z}
\hat{c}_i^\prime(z(t))=\hat{c}_i^\prime(z_0)+\sum_{i=0}^m \int_0^t \hat{c}_i^{\prime\prime}(z(\tau))u_i(\tau)\,d\tau.
\endeq
Noting that $y(t)=\hat{c}(z(t))$ and substituting \rref{eq:hat-c-prime-z} into \rref{eq:c-prime-z-again} gives
\begin{align*}
y(t)&=\hat{c}(z_0)+\sum_{i=0}^m  L_{V_i}\hat{c}(z_0)\int_0^tu_i(\tau)\,d\tau+ \\
&\hspace*{0.2in} \sum_{i_1,i_2=0}^m \int_0^t \int_0^{\tau_1} L_{V_{i_1}}\hat{c}_{i_2}(z(\tau_2))u_{i_2}(\tau_2)\,d\tau_2\, u_{i_1}(\tau_1)\,d\tau_1.
\end{align*}
Continuing in this way yields
\begin{align*}
y(t)&=\sum_{\eta\in X^\ast} L_{\mu(\eta)}\hat{c}(z_0)E_{\eta}[u](t) \\
&=\sum_{\eta\in X^\ast} \langle d,\eta\rangle E_{\eta}[u](t),
\end{align*}
which proves the theorem.
\endpr

\section{Networks of Chen--Fliess Series}
\label{sec:networks-Chen-Fliess-series}

In this section specific types of networks of Chen--Fliess series are
considered for which both Lemma~\ref{lem:well-defined-formal-representations} and
Theorem~\ref{th:generating-series-from-formal-representations} apply. To avoid a
barrage of indices, the component systems are assumed to be single-input, single-output.
There is, however, no technical reason for avoiding the multivariable case.
A variety of different configurations are possible. The following is perhaps the
simplest.

\begde
A set of $m$ single-input, single-output Chen--Fliess series mapping $u_i\mapsto y_i$ with generating series
$c_i\in\re_{LC}\langle\langle X_i\rangle\rangle$, where $X_i=\{x_0,x_i\}$,
and weighting matrix $M\in\re^{m\times m}$
is said to be \bfem{additively interconnected} if $u_i=v_i+\sum_{j=1}^mM_{ij}y_j$, $i=1,2,\ldots,m$.
\endde

In the following theorem, let $\mbf{e}_i\in\allseriesmLC$ denote the series with the $i$-th component series
being the monomial $\mbf{1}$, and the remaining components are the series having all coefficients equal to zero. In addition, given $c_j\in\allseriesLC$, define $\hat{c}_j=\mbf{1}\otimes\cdots\otimes\mbf{1}\otimes c_j\otimes\mbf{1}\cdots\otimes\mbf{1}\in\allseriesLCtensorm$, where $c_j$ appears in the $j$-th position.

\begth \label{th:additive-interconnections}
The input-output map $v\mapsto y$ of any additive interconnection of $m$ single-input, single-output Chen--Fliess series with generating
series $c_i\in\re_{LC}\langle\langle X_i\rangle\rangle$ has a well defined generating series $d\in\allseriesm$, where $d_j$ has the formal
representation $(\mu,\mbf{1}_m,\hat{c}_j)$ with $\mu$ defined in terms of the vector fields
\begdi
V_0(z)=
\left[
\begin{array}{c}
x_0z_1 \\
x_0z_2 \\
\vdots \\
x_0z_m
\end{array}
\right]+\diag(x_1z_1,\ldots,x_m z_m)M
\left[
\begin{array}{c}
\langle c_1,z_1\rangle \\
\langle c_2,z_2\rangle \\
\vdots \\
\langle c_m,z_m\rangle
\end{array}
\right],
\enddi
and $V_i(z)=x_iz_i\mbf{e}_i$ for $i=1,2,\ldots,m$.
\endth

\begpr
It is straightforward to show that the set of interconnected Chen--Fliess series constitutes an $m$ input, $m$ output system with
formal realization given by the vector fields as shown.
Therefore, the claim
follows directly from Lemma~\ref{lem:well-defined-formal-representations} and Theorem~\ref{th:generating-series-from-formal-representations} with
$\mu(x_i)=V_i$, $i=0,1,\ldots m$, $z_0=\mbf{1}_m$, and $\hat{c}_j\in\allseriesLCtensorm$.
\endpr

\begin{figure}[h]
\begin{center}
\includegraphics[scale=0.7]{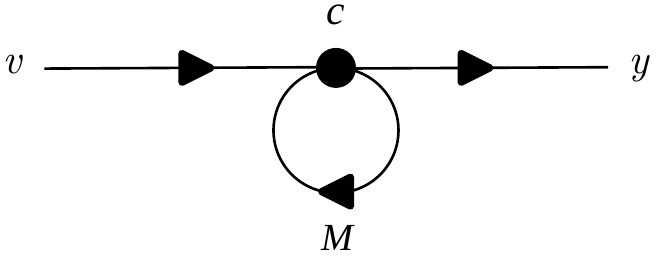}
\end{center}
\vspace*{-0.2in}
\caption{Single system additively interconnected}
\label{fig:one-node-system}
\end{figure}

\begex \label{ex:single-node-network} \rm
A single system additively interconnected with itself as shown in Figure~\ref{fig:one-node-system} would correspond to propositional output feedback, i.e.,
$u=v+My$ (dropping all subscripts). Thus, the corresponding representation is given by
\begdi
V_0(z)=(x_0+x_1M\langle c,z\rangle)z,\;\; V_1(z)=x_1z,
\enddi
$z_0=\mbf{1}_1=\mbf{1}$, and $\hat{c}=c$.
For a unity feedback system, i.e., $M=1$, applying \rref{eq:formal-representation} gives the following
generating series for the closed-loop system:
\begin{align*}
\langle d, \mbf{1}\rangle&=c(\mbf{1})=\langle c, \mbf{1}\rangle \\
\langle d, x_1\rangle&=L_{V_1}c(\mbf {1})=\langle c, x_1\rangle \\
\langle d, x_0\rangle&=L_{V_0}c(\mbf{1})=\langle c, x_0\rangle + \langle c, x_1 \rangle\langle c, \mbf{1} \rangle \\
\langle d, x_1^2\rangle&=L_{V_1}L_{V_1}c(\mbf{1})=\langle c, x_1^2\rangle \\
\langle d, x_0x_1\rangle&=L_{V_1}L_{V_0}c(\mbf{1})=\langle c, x_0x_1\rangle + \langle c, x_1 \rangle \langle c, x_1 \rangle +
\langle c, x_1^2 \rangle \langle c, \mbf{1} \rangle \\
\langle d, x_1x_0\rangle&=L_{V_0}L_{V_1}c(\mbf{1})= \langle c, x_1x_0 \rangle+\langle c, x_1^2\rangle \langle c, \mbf{1}\rangle \\
\langle d, x_0^2\rangle&=L_{V_0}L_{V_0}c(\mbf{1})= \langle c, x_0^2 \rangle+\langle c, x_1 \rangle \langle c, x_0 \rangle+ \langle c, x_1x_0\rangle\langle c, \mbf{1}\rangle+ \\
&\hspace*{0.2in}\langle c, x_0x_1\rangle\langle c, \mbf{1}\rangle +
\langle c, x_1\rangle\langle c, x_1\rangle\langle c, \mbf{1}\rangle+\langle c, x_1^2\rangle\langle c, \mbf{1}\rangle\langle c, \mbf{1}\rangle \\
&\hspace*{0.08in}\vdots
\end{align*}
These expressions are consistent with those in \cite{Gray-etal_SCL14}, where $d=S(-c)$, and $S$ is the
antipode of the output feedback Hopf algebra.
\endex

\begin{figure}[h]
\begin{center}
\includegraphics[scale=0.7]{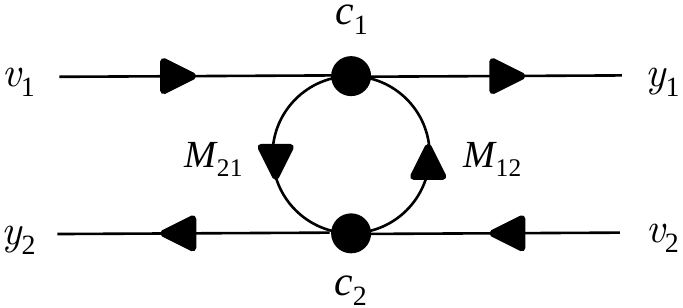}
\end{center}
\vspace*{-0.2in}
\caption{Two systems additively interconnected}
\label{fig:two-node-system}
\end{figure}

\begex \rm
Consider two additively interconnected systems as shown in Figure~\ref{fig:two-node-system}, where $M_{ij}=0$ when $i=j$.
Setting $M_{ij}=1$ for $i\neq j$ gives
a representation of $d_j$ specified by
\begdi
V_0(z)=
\left[
\begin{array}{c}
(x_0+x_1\langle c_2,z_2\rangle)z_1 \\
(x_0+x_2\langle c_1,z_1\rangle)z_2
\end{array}
\right],\;\; V_i(z)=x_iz_i \mbf{e}_i,\;\; i=1,2,
\enddi
$z_0=\mbf{1}_2$, and $\hat{c}_j$. For example,
the generating series $d_1$ for the mapping $v\mapsto y_1$ is:
\begin{align*}
\langle d_1, \mbf{1}\rangle&=\hat{c}_1(\mbf{1}_2)=\langle c_1,\mbf{1}\rangle \\
\langle d_1,x_1\rangle&=L_{V_1}\hat{c}_1(\mbf{1}_2)=\langle c_1,x_1\rangle \\
\langle d_1,x_2\rangle&=L_{V_2}\hat{c}_1(\mbf{1}_2)=0\\
\langle d_1,x_0\rangle&=L_{V_0}\hat{c}_1(\mbf{1}_2)=\langle c_1,x_0 \rangle+\langle c_1,x_1\rangle\langle c_2,\mbf{1} \rangle \\
\langle d_1,x_1^2\rangle&=L_{V_1}L_{V_1}\hat{c}_1(\mbf{1}_2)=\langle c_1,x_1^2\rangle \\
\langle d_1,x_1x_2\rangle&=L_{V_2}L_{V_1}\hat{c}_1(\mbf{1}_2)=0 \\
\langle d_1,x_2x_1\rangle&=L_{V_1}L_{V_2}\hat{c}_1(\mbf{1}_2)=0 \\
\langle d_1,x_2^2\rangle&=L_{V_2}L_{V_2}\hat{c}_1(\mbf{1}_2)=0 \\
\langle d_1,x_1x_0\rangle&=L_{V_0}L_{V_1}\hat{c}_1(\mbf{1}_2)=\langle c_1,x_1x_0 \rangle +\langle c_1,x_1^2\rangle\langle c_2,\mbf{1} \rangle \\
\langle d_1,x_0x_1\rangle&=L_{V_1}L_{V_0}\hat{c}_1(\mbf{1}_2)=\langle c_1,x_0x_1 \rangle +\langle c_1,x_1^2\rangle\langle c_2,\mbf{1} \rangle \\
&\hspace*{0.08in}\vdots
\end{align*}
and similarly for $d_{2}$ corresponding to the map $v\mapsto y_2$.
Unlike the first example, for networks with more than one system,
there is at present no known alterative algebraic method against which to compare all of these results.
Coefficient $\langle d_1,\eta\rangle$, where $\eta\in X_j^\ast$ and $j=1,2$ can be determined using the feedback product as described in \cite{Gray-etal_SCL14},
but {\em mixed} coefficients like $\langle d_1,x_1x_2\rangle$ can not.
\endex

\begin{figure}[h]
\begin{center}
\includegraphics[scale=0.7]{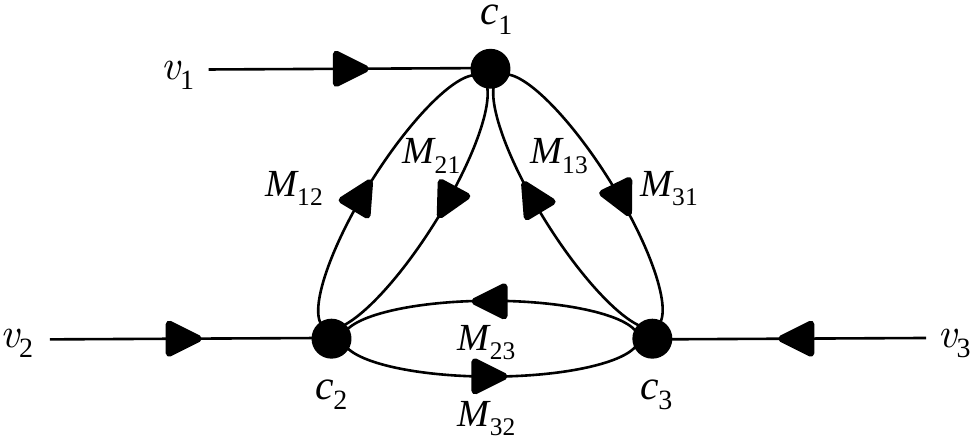}
\end{center}
\vspace*{-0.2in}
\caption{Three systems additively interconnected}
\label{fig:three-node-system}
\end{figure}

\begex \rm
Consider three additively interconnected systems as shown in Figure~\ref{fig:three-node-system}, where again $M_{ij}=0$ when $i=j$, and the
output branches have been suppressed.
For the case where $M_{ij}=1$ when $i\neq j$, a representation of $d_j$ is given by
\begin{align*}
V_0(z)&=
\left[
\begin{array}{c}
(x_0+x_1\langle c_2,z_2\rangle+x_1\langle c_3,z_3\rangle)z_1 \\
(x_0+x_2\langle c_1,z_1\rangle+x_2\langle c_3,z_3\rangle)z_2 \\
(x_0+x_3\langle c_1,z_1\rangle+x_3\langle c_2,z_2\rangle)z_3
\end{array}
\right] \\
V_i(z)&=x_iz_i\mbf{e}_i,\;\; i=1,2,3,
\end{align*}
$z_0=\mbf{1}_3$, and $\hat{c}_j$. For example,
the generating series $d_1$ for the mapping $v\mapsto y_1$ is:
\begin{align*}
\langle d_1,\mbf{1}\rangle&=\hat{c}_1(\mbf{1}_3)=\langle c_1,\mbf{1}\rangle \\
\langle d_1,x_1\rangle&=L_{V_1}\hat{c}_1(\mbf{1}_3)=\langle c_1,x_1\rangle \\
\langle d_1,x_2\rangle&=L_{V_2}\hat{c}_1(\mbf{1}_3)=0\\
\langle d_1,x_3\rangle&=L_{V_3}\hat{c}_1(\mbf{1}_3)=0\\
\langle d_1,x_0\rangle&=L_{V_0}\hat{c}_1(\mbf{1}_3)=\langle c_1,x_0 \rangle+\langle c_1,x_1\rangle\langle c_2,\mbf{1} \rangle+\langle c_1,x_1\rangle\langle c_3,\mbf{1} \rangle \\
\langle d_1,x_1^2\rangle&=L_{V_1}L_{V_1}\hat{c}_1(\mbf{1}_3)=\langle c_1,x_1^2\rangle \\
\langle d_1,x_1x_2\rangle&=L_{V_2}L_{V_1}\hat{c}_1(\mbf{1}_3)=0 \\
\langle d_1,x_1x_3\rangle&=L_{V_3}L_{V_1}\hat{c}_1(\mbf{1}_3)=0 \\
\langle d_1,x_1x_0\rangle&=L_{V_0}L_{V_1}\hat{c}_1(\mbf{1}_3)=\langle c_1,x_1x_0 \rangle +\langle c_1,x_1^2\rangle\langle c_2,\mbf{1} \rangle+
\langle c_1,x_1^2\rangle\langle c_3,\mbf{1} \rangle \\
\langle d_1,x_0x_1\rangle&=L_{V_1}L_{V_0}\hat{c}_1(\mbf{1}_3)=\langle c_1,x_0x_1 \rangle +\langle c_1,x_1^2\rangle\langle c_2,\mbf{1} \rangle+
\langle c_1,x_1^2\rangle\langle c_3,\mbf{1} \rangle \\
&\hspace*{0.08in}\vdots
\end{align*}
and similarly for $d_i$ corresponding to the map $v\mapsto y_i$, $i=2,3$.
\endex

Free from the bonds of linearity, other types of interconnections are also possible as considered next.

\begde
A set of $m$ single-input, single-output Chen--Fliess series mapping $u_i\mapsto y_i$ with generating series
$c_i\in\re_{LC}\langle\langle X_i\rangle\rangle$, where $X_i=\{x_0,x_i\}$, and weighting matrix $M\in\re^{m\times m}$
is said to be \bfem{multiplicatively interconnected} if $u_i=v_i\prod_{j=1}^mM_{ij}y_j$, $i=1,2,\ldots,m$.
\endde

\begth \label{th:multiplicative-interconnections}
Every input-output map $v\mapsto y$ of any multiplicative interconnection of $m$ single-input, single-output Chen--Fliess series with generating
series $c_i\in\re_{LC}\langle\langle X_i\rangle\rangle$ has a well defined generating series $d\in\allseriesm$, where $d_j$ has the formal
representation $(\mu,\mbf{1}_m,\hat{c}_j)$ with $\mu$ defined in terms of the vector fields
\begdi
V_0(z)=
\left[
\begin{array}{c}
x_0z_1 \\
x_0z_2 \\
\vdots \\
x_0z_m
\end{array}
\right],\;\;
V_i(z)=x_i\prod_{j=1}^m M_{ij}\langle c_j,z_j\rangle z_i\mbf{e}_i.
\enddi
\endth

\begpr
The proof is perfectly analogous to that of Theorem~\ref{th:additive-interconnections}.
\endpr

\begex \rm
Reconsider the single system network in Example~\ref{ex:single-node-network} except now multiplicatively interconnected,
that is, $u=vMy$ (again dropping all subscripts). The corresponding representation is given by
\begdi
V_0(z)=x_0z,\;\; V_1(z)=x_1M\langle c,z\rangle z,
\enddi
$z_0=\mbf{1}$, and $\hat{c}=c$.
Setting $M=1$ and applying \rref{eq:formal-representation} gives the following
generating series for the closed-loop system:
\begin{align*}
\langle d, \mbf{1}\rangle&=c(\mbf{1})=\langle c,\mbf{1}\rangle \\
\langle d, x_1\rangle&=L_{V_1}c(\mbf {1})=\langle c,x_1\rangle \langle c,\mbf{1}\rangle \\
\langle d, x_0\rangle&=L_{V_0}c(\mbf{1})=\langle c,x_0\rangle\\
\langle d, x_1^2\rangle&=L_{V_1}L_{V_1}c(\mbf{1})=\langle c,x_1^2\rangle\langle c,\mbf{1}\rangle\langle c,\mbf{1}\rangle+
\langle c,x_1\rangle\langle c,x_1\rangle\langle c,\mbf{1}\rangle \\
\langle d, x_0x_1\rangle&=L_{V_1}L_{V_0}c(\mbf{1})=\langle c,x_0x_1\rangle\langle c,\mbf{1}\rangle \\
\langle d, x_1x_0\rangle&=L_{V_0}L_{V_1}c(\mbf{1})=\langle c,x_1x_0\rangle\langle c,\mbf{1}\rangle+\langle c,x_1\rangle\langle c,x_0\rangle  \\
\langle d, x_1^3\rangle&=L_{V_1}L_{V_1}L_{V_1}c(\mbf{1})=\langle c,x_1^3\rangle\langle c,\mbf{1}\rangle\langle c,\mbf{1}\rangle\langle c,\mbf{1}\rangle+
4\langle c,x_1^2\rangle\langle c,x_1\rangle\langle c,\mbf{1}\rangle\langle c,\mbf{1}\rangle+ \\
&\hspace*{0.2in} \langle c,x_1\rangle\langle c,x_1\rangle\langle c,x_1\rangle\langle c,\mbf{1}\rangle \\
\langle d, x_0^2\rangle&=L_{V_0}L_{V_0}c(\mbf{1})=\langle c,x_0^2\rangle \\
&\hspace*{0.08in}\vdots
\end{align*}
Consider the particular case where $c=\sum_{k\geq 0}k!\,x_1^k$. Applying the
formulas above gives the closed-loop generating series
\begdi
d=1+x_1+3x_1^2+15x_1^3+\cdots,
\enddi
which is consistent with what was computed in \cite[Example~4.10]{Gray-Ebrahimi-Fard_SIAM17} using the antipode of the output affine feedback Hopf algebra.
\endex

\section{Conclusions and Future Work}

Using the concept of a formal realization and a formal representation,
it was shown that any additive or multiplicative interconnection of a set of
convergent single-input, single-output Chen--Fliess series always has a Chen--Fliess series representation whose
generating series can be computed explicitly
in terms of iterated formal Lie derivatives. This of course does not exhaust the list of possible network topologies for which this method is suitable.
For example, there can be mixtures of additive and multiplicative nodes in a given
network. There is also no technical barrier to applying the methodology in the full multivariable setting. Finally, the issue of convergence of the
network's generating series needs to be addressed in every case.

\section*{Acknowledgments}
The first author was supported by the National Science
Foundation under grant CMMI-1839378. The second author
was supported by the Research Council of Norway through project 302831
Computational Dynamics and Stochastics on Manifolds (CODYSMA).

\end{document}